\input amssym.tex
\input epsf
\epsfclipon


\magnification=\magstephalf
\hsize=14.0 true cm
\vsize=19 true cm
\hoffset=1.0 true cm
\voffset=2.0 true cm

\abovedisplayskip=12pt plus 3pt minus 3pt
\belowdisplayskip=12pt plus 3pt minus 3pt
\parindent=1.0em


\font\sixrm=cmr6
\font\eightrm=cmr8
\font\ninerm=cmr9

\font\sixi=cmmi6
\font\eighti=cmmi8
\font\ninei=cmmi9

\font\sixsy=cmsy6
\font\eightsy=cmsy8
\font\ninesy=cmsy9

\font\sixbf=cmbx6
\font\eightbf=cmbx8
\font\ninebf=cmbx9

\font\eightit=cmti8
\font\nineit=cmti9

\font\eightsl=cmsl8
\font\ninesl=cmsl9

\font\sixss=cmss8 at 8 true pt
\font\sevenss=cmss9 at 9 true pt
\font\eightss=cmss8
\font\niness=cmss9
\font\tenss=cmss10

 at 12 true pt
 at 12 true pt
\font\bigrm=cmr10 at 12 true pt
 at 12 true pt
 at 12 true pt

 at 16 true pt
 at 16 true pt
 at 16 true pt
 at 16 true pt
\font\Bigrm=cmr12 at 16 true pt
 at 16 true pt
 at 16 true pt

\catcode`@=11
\newfam\ssfam

\def\tenpoint{\def\rm{\fam0\tenrm}%
    \textfont0=\tenrm \scriptfont0=\sevenrm \scriptscriptfont0=\fiverm
    \textfont1=\teni  \scriptfont1=\seveni  \scriptscriptfont1=\fivei
    \textfont2=\tensy \scriptfont2=\sevensy \scriptscriptfont2=\fivesy
    \textfont3=\tenex \scriptfont3=\tenex   \scriptscriptfont3=\tenex
    \textfont\itfam=\tenit                  \def\it{\fam\itfam\tenit}%
    \textfont\slfam=\tensl                  \def\sl{\fam\slfam\tensl}%
    \textfont\bffam=\tenbf \scriptfont\bffam=\sevenbf
    \scriptscriptfont\bffam=\fivebf
                                            \def\bf{\fam\bffam\tenbf}%
    \textfont\ssfam=\tenss \scriptfont\ssfam=\sevenss
    \scriptscriptfont\ssfam=\sevenss
                                            \def\ss{\fam\ssfam\tenss}%
    \normalbaselineskip=13pt
    \setbox\strutbox=\hbox{\vrule height8.5pt depth3.5pt width0pt}%
    \let\big=\tenbig
    \normalbaselines\rm}

\def\ninepoint{\def\rm{\fam0\ninerm}%
    \textfont0=\ninerm      \scriptfont0=\sixrm
                            \scriptscriptfont0=\fiverm
    \textfont1=\ninei       \scriptfont1=\sixi
                            \scriptscriptfont1=\fivei
    \textfont2=\ninesy      \scriptfont2=\sixsy
                            \scriptscriptfont2=\fivesy
    \textfont3=\tenex       \scriptfont3=\tenex
                            \scriptscriptfont3=\tenex
    \textfont\itfam=\nineit \def\it{\fam\itfam\nineit}%
    \textfont\slfam=\ninesl \def\sl{\fam\slfam\ninesl}%
    \textfont\bffam=\ninebf \scriptfont\bffam=\sixbf
                            \scriptscriptfont\bffam=\fivebf
                            \def\bf{\fam\bffam\ninebf}%
    \textfont\ssfam=\niness \scriptfont\ssfam=\sixss
                            \scriptscriptfont\ssfam=\sixss
                            \def\ss{\fam\ssfam\niness}%
    \normalbaselineskip=12pt
    \setbox\strutbox=\hbox{\vrule height8.0pt depth3.0pt width0pt}%
    \let\big=\ninebig
    \normalbaselines\rm}

\def\eightpoint{\def\rm{\fam0\eightrm}%
    \textfont0=\eightrm      \scriptfont0=\sixrm
                             \scriptscriptfont0=\fiverm
    \textfont1=\eighti       \scriptfont1=\sixi
                             \scriptscriptfont1=\fivei
    \textfont2=\eightsy      \scriptfont2=\sixsy
                             \scriptscriptfont2=\fivesy
    \textfont3=\tenex        \scriptfont3=\tenex
                             \scriptscriptfont3=\tenex
    \textfont\itfam=\eightit \def\it{\fam\itfam\eightit}%
    \textfont\slfam=\eightsl \def\sl{\fam\slfam\eightsl}%
    \textfont\bffam=\eightbf \scriptfont\bffam=\sixbf
                             \scriptscriptfont\bffam=\fivebf
                             \def\bf{\fam\bffam\eightbf}%
    \textfont\ssfam=\eightss \scriptfont\ssfam=\sixss
                             \scriptscriptfont\ssfam=\sixss
                             \def\ss{\fam\ssfam\eightss}%
    \normalbaselineskip=10pt
    \setbox\strutbox=\hbox{\vrule height7.0pt depth2.0pt width0pt}%
    \let\big=\eightbig
    \normalbaselines\rm}

\def\tenbig#1{{\hbox{$\left#1\vbox to8.5pt{}\right.\n@space$}}}
\def\ninebig#1{{\hbox{$\textfont0=\tenrm\textfont2=\tensy
                       \left#1\vbox to7.25pt{}\right.\n@space$}}}
\def\eightbig#1{{\hbox{$\textfont0=\ninerm\textfont2=\ninesy
                       \left#1\vbox to6.5pt{}\right.\n@space$}}}

\font\sectionfont=cmbx10
\font\subsectionfont=cmti10

\def\figurecaptionfont{\ninepoint}
\def\tablecaptionfont{\ninepoint}
\def\footnotefont{\eightpoint}


\newcount\equationno
\newcount\bibitemno
\newcount\figureno
\newcount\tableno

\equationno=0
\bibitemno=0
\figureno=0
\tableno=0


\footline={\ifnum\pageno=0{\hfil}\else
{\hss\rm\the\pageno\hss}\fi}


\def\section #1. #2 \par
{\vskip0pt plus .10\vsize\penalty-100 \vskip0pt plus-.10\vsize
\vskip 1.6 true cm plus 0.2 true cm minus 0.2 true cm
\global\def\equationlabel{#1}
\global\equationno=0
\leftline{\sectionfont #1. #2}\par
\immediate\write\terminal{Section #1. #2}
\vskip 0.7 true cm plus 0.1 true cm minus 0.1 true cm
\noindent}


\def\subsection #1 \par
{\vskip0pt plus 0.8 true cm\penalty-50 \vskip0pt plus-0.8 true cm
\vskip2.5ex plus 0.1ex minus 0.1ex
\leftline{\subsectionfont #1}\par
\immediate\write\terminal{Subsection #1}
\vskip1.0ex plus 0.1ex minus 0.1ex
\noindent}


\def\appendix #1. #2 \par
{\vskip0pt plus .20\vsize\penalty-100 \vskip0pt plus-.20\vsize
\vskip 1.6 true cm plus 0.2 true cm minus 0.2 true cm
\global\def\equationlabel{\hbox{\rm#1}}
\global\equationno=0
\leftline{\sectionfont Appendix #1. #2}\par
\immediate\write\terminal{Appendix #1. #2}
\vskip 0.7 true cm plus 0.1 true cm minus 0.1 true cm
\noindent}



\def\equation#1{$$\displaylines{\qquad #1}$$}
\def\enum{\global\advance\equationno by 1
\hfill\llap{{\rm(\equationlabel.\the\equationno)}}}
\def\noenum{\hfill}
\def\next#1{\cr\noalign{\vskip#1}\qquad}


\def\ifundefined#1{\expandafter\ifx\csname#1\endcsname\relax}

\def\ref#1{\ifundefined{#1}?\immediate\write\terminal{unknown reference
on page \the\pageno}\else\csname#1\endcsname\fi}

\newwrite\terminal
\newwrite\bibitemlist

\def\bibitem#1#2\par{\global\advance\bibitemno by 1
\immediate\write\bibitemlist{\string\def
\expandafter\string\csname#1\endcsname
{\the\bibitemno}}
\item{[\the\bibitemno]}#2\par}

\def\beginbibliography{
\vskip0pt plus .15\vsize\penalty-100 \vskip0pt plus-.15\vsize
\vskip 1.2 true cm plus 0.2 true cm minus 0.2 true cm
\leftline{\sectionfont References}\par
\immediate\write\terminal{References}
\immediate\openout\bibitemlist=biblist
\frenchspacing\parindent=1.6em
\vskip 0.5 true cm plus 0.1 true cm minus 0.1 true cm}

\def\endbibliography{
\immediate\closeout\bibitemlist
\nonfrenchspacing\parindent=1.0em}


\def\figurecaption#1{\global\advance\figureno by 1
\narrower\figurecaptionfont
Fig.~\the\figureno. #1}

\def\tablecaption#1{\global\advance\tableno by 1
\vbox to 0.5 true cm { }
\centerline{\tablecaptionfont%
Table~\the\tableno. #1}
\vskip-0.4 true cm}

\def\thintablerule{\hrule height0.4pt}

\tenpoint

\immediate\openin\bibitemlist=biblist
\ifeof\bibitemlist\immediate\closein\bibitemlist
\else\immediate\closein\bibitemlist
\input biblist \fi


\def\thismonth{\ifcase\month\or
January\or February\or March\or April\or May\or June\or
July\or August\or September\or October\or November\or December\fi}



\def\rmd{{\rm d}}

\def\rme{{\rm e}}
\def\rmO{{\rm O}}


\def\Re{{\rm Re}\,}


\def\proof{\noindent{\sl Proof:}\kern0.6em}

\def\frac#1#2{\hbox{$#1\over#2$}}
\def\dual{\mathstrut^*\kern-0.1em}

\def\lvec#1{\setbox0=\hbox{$#1$}
    \setbox1=\hbox{$\scriptstyle\leftarrow$}
    #1\kern-\wd0\smash{
    \raise\ht0\hbox{$\raise1pt\hbox{$\scriptstyle\leftarrow$}$}}
    \kern-\wd1\kern\wd0}
\def\rvec#1{\setbox0=\hbox{$#1$}
    \setbox1=\hbox{$\scriptstyle\rightarrow$}
    #1\kern-\wd0\smash{
    \raise\ht0\hbox{$\raise1pt\hbox{$\scriptstyle\rightarrow$}$}}
    \kern-\wd1\kern\wd0}


\def\nabstar#1{{\nabla\kern0.5pt\smash{\raise 4.5pt\hbox{$\ast$}}
               \kern-5.5pt_{#1}}}

\def\drvstar#1{{\partial\kern0.5pt\smash{\raise 4.5pt\hbox{$\ast$}}
               \kern-6.0pt_{#1}}}

\def\ldrvstar#1{{\lvec{\,\partial}\kern-0.5pt\smash{\raise 4.5pt\hbox{$\ast$}}
               \kern-5.0pt_{#1}}}





\def\dirac#1{\gamma_{#1}}
\def\diracstar#1#2{
    \setbox0=\hbox{$\gamma$}\setbox1=\hbox{$\gamma_{#1}$}
    \gamma_{#1}\kern-\wd1\kern\wd0
    \smash{\raise4.5pt\hbox{$\scriptstyle#2$}}}


\def\SUthree{{\rm SU(3)}}

\def\Ad{{\rm Ad}\kern0.1em}


\def\Sg{S_{\rm G}}

\def\Schi{S_{\chi}}
\def\B{\Lambda}
\def\Bs{\B^{\kern-0.5pt\ast}}
\def\dB{\partial\B}
\def\dBs{\partial\Bs}
\def\QB{Q_{\B}}
\def\QBs{Q_{\Bs}}
\def\QdB{Q_{\dB}}
\def\QdBs{Q_{\dBs}}
\def\thetaB{\theta_{\B}}
\def\thetaBs{\theta_{\Bs}}
\def\PrdB{\theta_{\dB}}
\def\rhoB{\varrho_{\B}}
\vbox{\vskip1.0cm}
\rightline{CERN-TH/2003-088}

\vskip 2.0cm 
\centerline{\Bigrm Lattice QCD and the Schwarz alternating procedure}
\vskip 0.6 true cm
\centerline{\bigrm Martin L\"uscher}
\vskip1ex
\centerline{\it CERN, Theory Division}
\centerline{\it CH-1211 Geneva 23, Switzerland}
\vskip 0.8 true cm
\thintablerule
\vskip 2.0ex
\ninepoint
\leftline{\bf Abstract}
\vskip 1.0ex\noindent
A numerical simulation algorithm for lattice QCD is described,
in which the short- and long-distance effects of the sea 
quarks are treated separately. The algorithm can be regarded,
to some extent,
as an implementation at the quantum level
of the classical Schwarz alternating procedure 
for the solution of elliptic partial differential equations.
No numerical tests are reported here, but 
theoretical arguments suggest that the algorithm
should work well also at small quark masses.

\vskip 2.0ex
\thintablerule

\tenpoint


\section 1. Introduction

The simulation algorithms for (unquenched) lattice QCD 
that are currently in use rapidly become inefficient
on large lattices and at small quark masses, where 
the effects of spontaneous chiral symmetry breaking set in.
In the case of the
HMC [\ref{HMC}], the PHMC [\ref{PHMCI},\ref{PHMCII}] 
and the Multiboson [\ref{MB}] algorithms,
the principal technical difficulty derives from the fact
that the light quark masses have to be scaled
proportionally to the square of the pion mass in the chiral limit.
The lattice Dirac operator is then increasingly ill-conditioned,
which affects all these algorithms in a similar and rather direct way,
because they all start from a {\it global pseudo-fermion representation} of 
the quark determinant that involves an exact (or an accurate
approximate) inversion of the Dirac operator.

In the present paper a simulation algorithm is proposed that 
exploits the underlying local structure of the theory and that
may be expected to scale in a more favourable way in the chiral regime.
The general strategy is closely related
to the alternating procedure that was invented by the
mathematician Schwarz in the 19th century
to establish the existence of the solution of the Dirichlet
problem 
\equation{
  \left.\Delta f(x)\right|_{x\in\Omega}=0,
  \qquad
  \left.f(x)\right|_{x\in\partial\Omega}=g(x),
  \enum
}
on arbitrary bounded domains $\Omega$ in the plane [\ref{Schwarz}]
(for an introduction to the subject in the context of
discretized partial differential equations
see ref.~[\ref{Saad}], for example).
Very briefly this method obtains the solution iteratively
by dividing $\Omega$ into a set of overlapping subdomains
and by solving the Dirichlet problem on these, in each step of the iteration,
with boundary values determined from the current approximation
to the solution.

The design of simulation algorithms for lattice QCD that 
operate on overlapping blocks of lattice points is
non-trivial, 
however, because the global correctness of the simulation
must be guaranteed. 
In principle the problem can be solved
using stochastic acceptance--rejection steps
similar to those previously considered by Hasenbusch 
[\ref{Hasenbusch}] (see also refs.~[\ref{FrezzottiEtAl}--\ref{PSD}]).
The key question is then
whether these correction steps can be implemented so that 
high acceptance rates are achieved, and
a significant part of the present paper is therefore devoted
to this issue.

\section 2. General form of the algorithm

In this section the proposed algorithm is described in outline.
Most technical details are deferred to the later sections
and important improvements (such as preconditioning) are omitted
in order to keep the presentation as simple as possible.
Some algorithm theory, as summarized in appendix A, 
is nevertheless required to be able
to understand the procedure.

\subsection 2.1 Preliminaries

Although the algorithm is more generally applicable,
only the standard Wilson formulation [\ref{Wilson}] of lattice 
QCD will be considered here
(optionally including $\rmO(a)$ improvement
[\ref{SW},\ref{OaImp}]) with  
a doublet of mass-degenerate quarks.
The lattice spacing is set to unity for convenience and 
the SU(3) link variables are
denoted by $U(x,\mu)$ as usual.

After integration over the quark fields, the
probability distribution
of the gauge field reads
\equation{
  P[U]={1\over\cal Z}\,\rme^{-\Sg}\det Q^2,
  \qquad
  Q\equiv\dirac{5}(D_{\rm w}+m_0),
  \enum
}
where 
$\cal Z$ denotes the partition function,
$\Sg$ the plaquette action,
$D_{\rm w}$ the Wilson--Dirac operator
and $m_0$ the bare quark mass.

\subsection 2.2 Alternating procedure

Following the classical Schwarz procedure, the proposed algorithm visits
rectangular blocks of lattice points according to some scheme
and updates the link variables residing there.
The blocks can have arbitrary sizes in principle, and their
position may be chosen randomly, for example,
so that all link variables are treated equally.
However, as will become clear later, 
the algorithm is designed to perform particularly well if
the blocks are small in physical units, 
i.e.~if their edges are less than about $1$~fm long.

On each block $\B$ that is visited in the course of this process,
changes in the gauge field on $\B$ are proposed that 
satisfy detailed balance with respect to the distribution
\equation{
  P_{\B}[U]={1\over{\cal Z}_{\B}}\,\rme^{-\Sg}
  \det(\QB+\QBs)^2.
  \enum
}
A stochastic acceptance--rejection step then needs to be 
applied to correct for the 
difference between this distribution and the exact distribution
(2.1).
The
operators $\QB$ and $\QBs$ that appear here
coincide with $Q$, except that they act
on Dirac fields defined on the block $\B$ and its complement $\Bs$,
respectively,
with Dirichlet boundary conditions.
At the level of the fermion action, a complete decoupling of 
the quark fields inside and outside
the block is thus achieved. In particular, 
the proposals for the link variables
residing in $\B$ can be generated locally using a block version 
of the HMC algorithm, for example.

The stochastic acceptance--rejection step, on the other hand, involves
an inversion of the Dirac operator $Q$ on the full lattice.
There is a fair amount of choice in the detailed implementation 
of this step, which can be exploited to reduce the influence
on the acceptance probability of the link variables far 
away from the block. Moreover, the suggested procedure
(which is explained in sect.~5) restricts
the pseudo-fermion field that needs
to be introduced at this stage to the boundary of the block.
At least to some extent,
the local character of the algorithm is thus preserved.

\subsection 2.3 Hierarchical structure

In the form described above the algorithm requires
a computational effort per block update that increases 
roughly proportionally 
to the lattice volume. 
By introducing a hierarchy of blocks such as the one shown 
in fig.~1,
one may, however, be able to do better than this.
Proposed changes of the 
link variables on the smallest blocks are then obtained
as before, and these are 
taken as proposed configurations 
on the next larger blocks, and so on, accepting or rejecting them so that
detailed balance is satisfied with respect to the 
associated distributions $P_{\B}[U]$.
Finally a {\it simultaneous}\/
global acceptance--rejection step is applied
to the surviving configurations.

\topinsert
\vbox{
\vskip0.0cm
\epsfxsize=5.0cm\leftline{\hfill\epsfbox{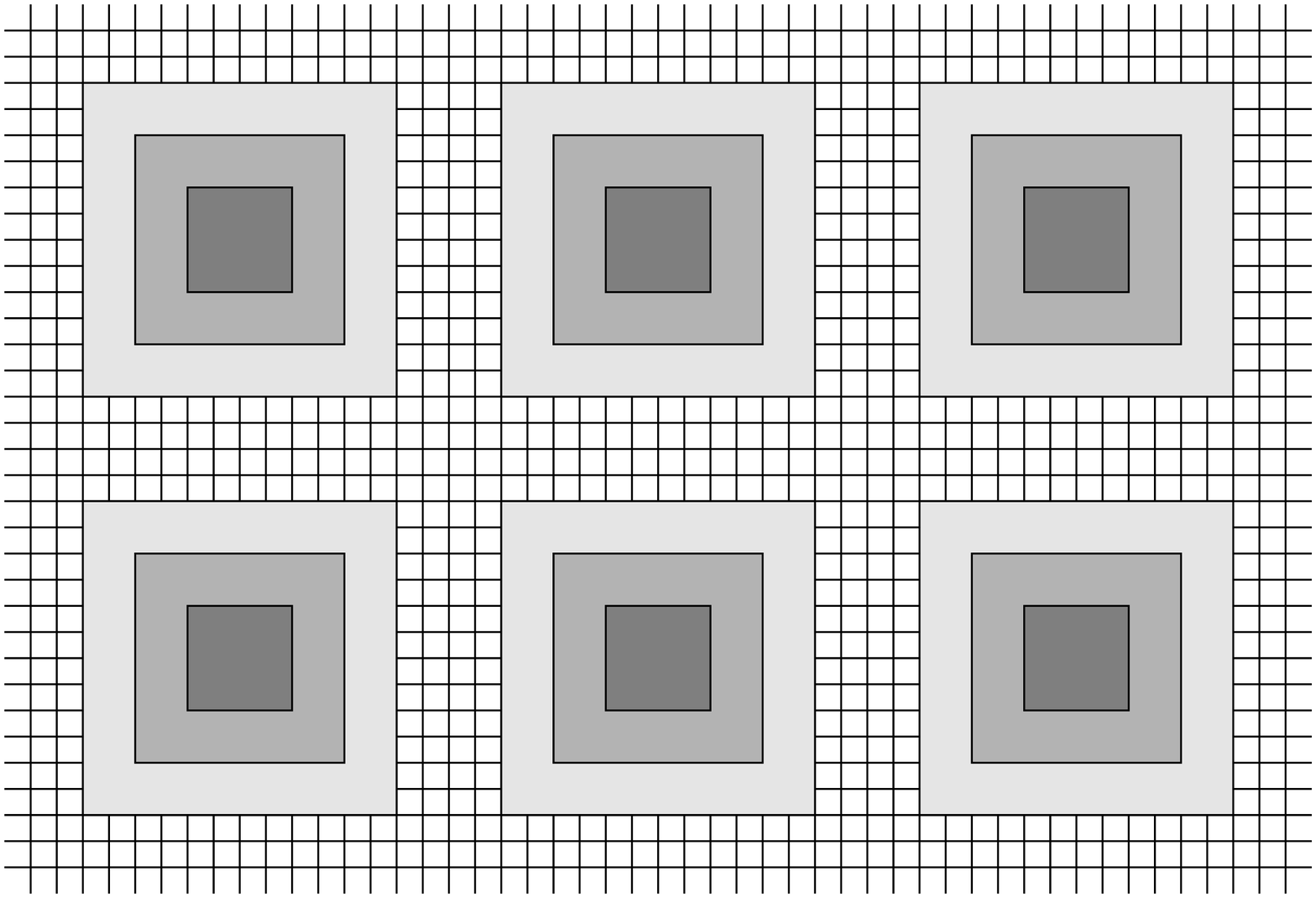}\hfill}
\vskip0.4cm
\figurecaption{%
Proposed updates of the link variables on small blocks 
(dark grey squares) can be filtered through
acceptance--rejection steps on increasingly larger blocks (light grey 
squares) before the global acceptance--rejection step is applied. 
}
\vskip0.1cm
}
\endinsert

This procedure is logically correct since 
non-overlapping blocks are decoupled from each other 
and can be updated in parallel. 
On the other hand, it will only work out in practice
if the configurations that have been 
accepted on the largest blocks have a high probability to be accepted
on the whole lattice. For the chosen implementation 
of the acceptance--rejection steps,
a theoretical argument 
will be given in sect.~5 that explains
why this should be so (under certain conditions).

\subsection 2.4 Low-mode reweighting

On physically small blocks $\B$, the Dirac operator $\QB$
that acts on the quark fields in $\B$ 
is not expected to become ill-conditioned
at small quark masses. The situation here
is actually similar to the one encountered in studies 
of the Schr\"odinger functional, where the boundary conditions
provide an infrared cutoff on both the gluon and the quark modes
[\ref{SchroedingerGauge},\ref{SchroedingerFermion}].
As a consequence the block simulation algorithm
should work well
even if the bare quark mass $m_0$ is set to the critical
mass $m_c$ [\ref{SchroedingerFullQCD}].

However, since the Wilson formulation of lattice QCD 
violates chiral symmetry,
the Dirac operator on large lattices
is not protected against the accidental presence of eigenvectors
with eigenvalues much smaller than $m_0-m_c$.
When a gauge field configuration is proposed where this happens,
it will only rarely pass the global acceptance--rejection step.
Such configurations are therefore sampled with low statistics
and this can easily lead to uncontrolled statistical fluctuations
if a quantity is considered that 
is sensitive to the low modes of the Dirac operator
(the propagator of the pseudo-scalar density, for example).

It may be possible to solve this problem by including 
the product 
\equation{
   W[U]=\prod_{k=1}^n\lambda_k
   \enum
}
of the first few eigenvalues $\lambda_1,\ldots,\lambda_n$ of $Q^2$
in the observables and 
the inverse factor $W[U]^{-1}$ 
in the global acceptance--rejection step.
The computational overhead for this modification 
may not be negligible, but it should be noted in this connection
that the eigenvalues do not need to be computed very accurately
(as long as a definite procedure is used that obtains the
eigenvalues independently of the previous gauge field configurations).
Moreover the associated 
approximate eigenvectors can help to accelerate the inversion
of the Dirac operator that is required in the acceptance--rejection
step [\ref{LowMode}].

\section 3. Domain decomposition

Let $\B$ be an arbitrary rectangular block of lattice points.
$\B$ and its 
complement $\Bs$ (the set of points not in $\B$) 
define a particular case of a domain decomposition of the lattice.
In the following paragraphs the aim is to introduce 
some basic notation related to this decomposition,
but the terminology is quite general and extends 
to the case where $\B$ is replaced by the union of 
a set of non-overlapping blocks.

\subsection 3.1 Boundary points

On the lattice it is important to distinguish between 
interior and exterior boundary points (see fig.~2).
The boundary values for 
the classical Dirichlet problem on $\B$, for example, should be specified on 
the set $\dB$ of all exterior boundary points
if the standard nearest-neighbour lattice laplacian is used,
while the set of interior
boundary points plays an analogous r\^ole from the point of view
of the complementary domain $\Bs$ and is therefore denoted
by $\dBs$. 

\topinsert
\vbox{
\vskip0.0cm
\epsfxsize=4.4cm\leftline{\hfill\epsfbox{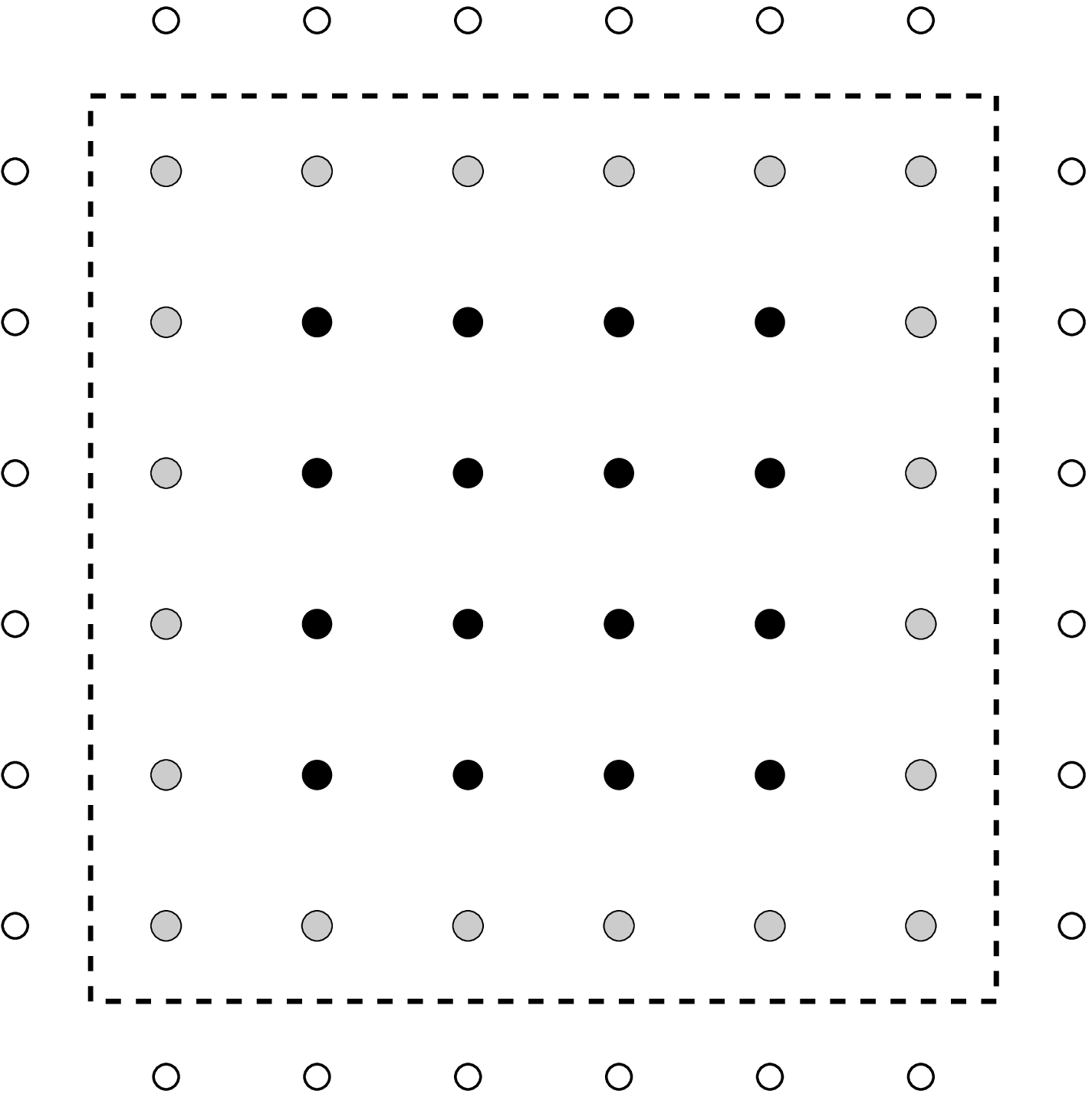}\hfill}
\vskip0.4cm
\figurecaption{%
Two-dimensional slice of a $6^4$ block (black and grey 
points inside the
dashed square). The grey and the open points represent
the interior and the exterior boundary points of the block
respectively.
}
\vskip0.2cm
}
\endinsert

A special feature of this geometry is that for each exterior boundary 
point there is a unique closest point on the interior boundary,
referred to as its partner point.
The converse is evidently not true, i.e.~different exterior
boundary points may have the same partner point.

\subsection 3.2 Decomposition of the Dirac operator

In position space the hermitian lattice Dirac operator is 
a sparse matrix that assumes the block-diagonal form
\equation{
  Q=\pmatrix{\QB\hfill\kern-0.5em &\QdB\hfill \cr
             \noalign{\vskip1ex}
             \QdBs\hfill\kern-0.5em &\QBs\hfill \cr}
  \enum
}
if the lattice points are ordered so that those in $\B$ come first.
The operator $\QB$, for example, acts on Dirac fields on $\B$
in the same way as $Q$, except that all terms
involving the
exterior boundary points are set to zero (which is equivalent to
imposing Dirichlet boundary conditions on $\dB$).

It is often convenient to consider
$\QB$, $\QBs$, $\QdB$ and $\QdBs$ to be operators 
in the space of Dirac fields that are defined on the whole lattice
rather than on $\B$ or $\Bs$ only.
The embedding is done in the obvious way by
padding with zeros. 
Equation (3.1) may then be written in the form
\equation{
   Q=\QB+\QBs+\QdB+\QdBs,
   \enum
}
and an equivalent expression for the determinant in eq.~(2.2) 
is $\det\QB^2\det\QBs^2$.
This notation is perhaps slightly abusive
but should not lead to any confusion since it is usually clear from 
the context which interpretation applies.  

The operator $\QdB$ and its hermitian conjugate $\QdBs$ connect the domains
$\B$ and $\Bs$ to each other.
Explicitly $\QdB$ is given by 
\equation{
  \QdB\psi(x)=-\dirac{5}\thetaB(x)
  \sum_{\mu=0}^3\Bigl\{\frac{1}{2}
  (1-\dirac{\mu})\thetaBs(x+\hat{\mu})U(x,\mu)\psi(x+\hat{\mu})
  \noenum
  \next{1.5ex}
  {\phantom{\QdB\psi(x)=}}\kern2.6em
  +\frac{1}{2}(1+\dirac{\mu})\thetaBs(x-\hat{\mu})U(x-\hat{\mu},\mu)^{-1}
  \psi(x-\hat{\mu})
  \Bigr\},
  \enum
}
where $\thetaB$ and $\thetaBs$ denote the characteristic functions 
of $\B$ and $\Bs$ respectively%
\kern1pt\footnote{$\dag$}{\footnotefont%
The normalization conventions and all unexplained notations are
as in ref.~[\ref{OaImp}].}.
This operator thus transports the Dirac spinors on 
the exterior boundary points to the partner points on the interior boundary,
while $\QdBs$ does the same in the opposite direction.

\subsection 3.3 Dirichlet problem \& space of boundary values

To be able to properly pose the Dirichlet boundary value problem
for the lattice Dirac operator on $\B$, the space of boundary 
values first needs to be specified. 
The situation is practically the same as in the case of 
the Schr\"odinger functional which was studied
in ref.~[\ref{SchroedingerFermion}].
It is useful in this connection to introduce
the projector-valued function
\equation{
   \PrdB(x)=\sum_{\mu=0}^3\thetaBs(x)\left\{
   \frac{1}{2}(1-\dirac{\mu})\thetaB(x-\hat{\mu})
   +
   \frac{1}{2}(1+\dirac{\mu})\thetaB(x+\hat{\mu})
   \right\}
   \enum
}
that will, in many respects, play the r\^ole of the characteristic function
of the exterior boundary.
The only non-vanishing terms on the right-hand side
of eq.~(3.4) are in fact those where $x$ is in $\dB$ and $x-\hat{\mu}$
or $x+\hat{\mu}$
its partner point,  and the projectors $\frac{1}{2}(1\pm\dirac{\mu})$
are precisely such that the identity
\equation{
  \QdB=\QdB\PrdB
  \enum
}
holds.

The Dirichlet boundary value problem for the lattice Dirac operator is now
to find a Dirac field $\psi(x)$ on 
$\B\cup\dB$ that satisfies
\equation{
   Q\psi(x)=0
   \quad\hbox{for all $ x\in\B$},
   \enum
   \next{2.0ex}
   \PrdB(x)\psi(x)=\psi(x)=\eta(x)
   \quad\hbox{for all $x\in\dB$},
   \enum
}
where $\eta=\PrdB\eta$ is any prescribed field on $\dB$.
Recalling the decomposition (3.1) of the Dirac operator, it 
is clear that $Q$ can be replaced by $\QB+\QdB$ in eq.~(3.6).
At all points in $\B$ the solution is then given by
\equation{
   \psi=-\QB^{-1}\QdB\eta,
   \enum
}
where use has been made of
the fact that $\QdB$ moves the field $\eta$ 
from the exterior to the interior boundary of the block.
In particular, the solution is uniquely determined 
by the specified boundary values 
(if $\QB$ is invertible).

\section 4. Block simulation algorithm

For the generation of proposed gauge field configurations on a given
block $\B$, the HMC and PHMC algorithms are both equally suitable.
Some relevant details are given in this section for 
the case of the HMC algorithm.
The correctness of the local form of this algorithm is easily
shown by noting that it matches the
general pattern outlined in subsect.~A.4.

\subsection 4.1 Active and spectator link variables

As already mentioned in sect.~2,
the proposed updates of the gauge field on $\B$ 
must satisfy detailed balance with respect to the
distribution (2.2), which is now written in the form 
\equation{ 
   P_{\B}[U]={1\over{\cal Z}_{\B}}\,\rme^{-\Sg}
  \det\QB^2\det\QBs^2.
  \enum
}
To achieve a complete decoupling from the surrounding lattice, 
only the link variables residing on a subset of links in $\B$,
the {\it active link variables}, should be changed.
In particular, if the links shown in fig.~3 are selected,
the factor $\det\QBs$ is guaranteed to be independent of the 
active link variables, independently of whether
the $\rmO(a)$-improved or the unimproved theory is considered.

\topinsert
\vbox{
\vskip0.0cm
\epsfxsize=4.4cm\leftline{\hfill\epsfbox{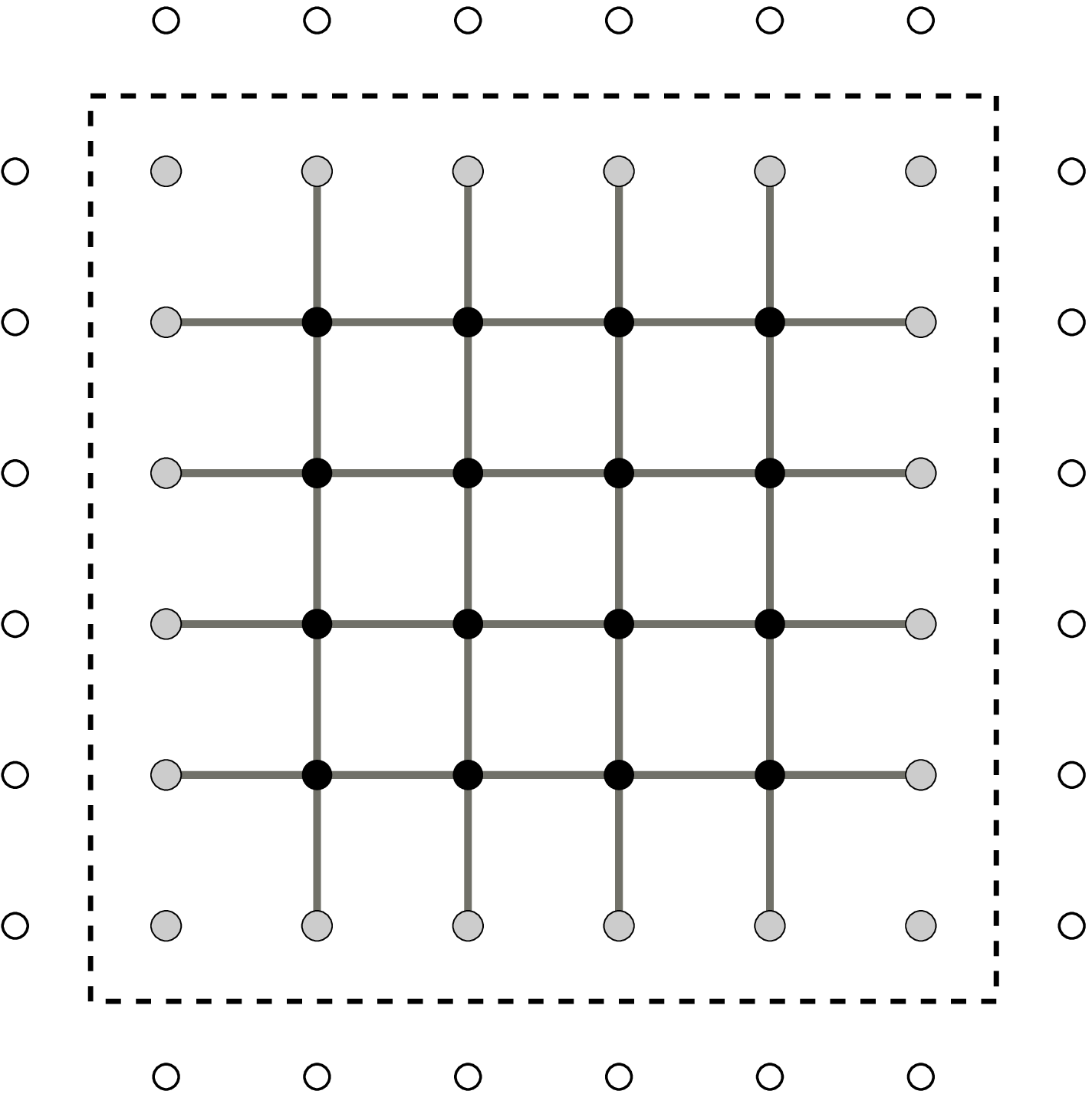}\hfill}
\vskip0.4cm
\figurecaption{%
The block update algorithm only changes the link variables on 
a subset of links in the chosen block (solid lines). In particular, the 
link variables on the interior boundary and those connecting
the interior to the exterior boundary are held fixed.
}
\vskip0.1cm
}
\endinsert

At this point the HMC algorithm can be set up as usual by 
introducing a pseudo-fermion field $\phi(x)$ on $\B$ and the 
canonical momenta $\Pi(x,\mu)$ of the active
link variables $U(x,\mu)$.
The total action of the system is then taken to be 
\equation{
  S=\Sg+\frac{1}{2}\left(\Pi,\Pi\right)+\left(\phi,\QB^{-2}\phi\right),
  \enum
}
where the brackets $(\cdot,\cdot)$ denote the natural scalar products
in the relevant spaces of fields. 
Note that the factor $\det\QBs$ can be ignored, since it
does not depend on the active link variables.
All inactive link variables actually play a spectator
r\^ole in the following, which 
is rather similar to what happens
in the case of the widely used simulation algorithms for
pure gauge theories, where the link variables are updated one
after the other.

\subsection 4.2 Variable-speed molecular dynamics

Following the standard procedure,
a new configuration of the active link variables is now 
generated by first choosing the pseudo-fermion field and 
the canonical momenta randomly with conditional probability
proportional to $\rme^{-S}$.
The momenta and the active link variables
are then evolved from their initial values
$\Pi,U$ to some final values $\Pi',U'$ by solving the molecular
dynamics equations
\equation{
  {\rmd\over\rmd\tau}\Pi(x,\mu)=-\gamma(x,\mu)F(x,\mu),
  \enum
  \next{2ex}
  {\rmd\over\rmd\tau}U(x,\mu)=\gamma(x,\mu)\Pi(x,\mu)U(x,\mu),
  \enum
}
from $\tau=0$ to $\tau=1$. 
As usual the force $F(x,\mu)$ takes values in the Lie algebra
of $\SUthree$ and is determined by the requirement that
\equation{
  \left(\omega,F\right)=
  \lim_{\epsilon\to0}\left\{{1\over\epsilon}
  \left(\left.S\right|_{U\to U_{\epsilon}}-S\right)\right\},
  \qquad
  U_{\epsilon}(x,\mu)\equiv\rme^{\epsilon\omega(x,\mu)}U(x,\mu),
  \enum
}
for all variations $\omega(x,\mu)$ of the active link variables. 
It should be emphasized that 
the computation of the force is an operation that 
is local to the block $\B$. In particular,
only the Dirac operator $\QB$ needs to be inverted.

The action and the integration measure are both preserved
in the course of the molecular dynamics evolution,
for any choice of the (real) speed factor $\gamma(x,\mu)$.
This factor is usually set to $1$,
but for reasons explained later,
it may be advisable, in the present context, to take a smaller
value on
the links close to the boundary of $\B$. The associated
link variables are then slowed down and tend to deviate less from
their initial values at the end of the evolution
(stochastic equations where different field components
are evolved at different speeds
have previously been considered
by Davies et al.~[\ref{DaviesEtAl}]).

It is, incidentally, straightforward to incorporate this modification
in the numerical integration scheme that is used to solve the
molecular dynamics equations. As in the case of the standard
HMC algorithm, an acceptance--rejection step is finally required
to correct for the integration error and thus to obtain
a transition probability that satisfies detailed balance
with respect to the distribution $P_{\B}[U]$.

\section 5. Global acceptance--rejection step

The gauge field configurations that are generated 
by applying the local algorithm described above
can be taken as proposed configurations
for the theory on a larger block or on the whole lattice.
Whether a proposed field is accepted or not is then 
decided by applying a stochastic acceptance--rejection step.
A particular implementation
of this step is now going to be discussed, first for the case
of the transition from a given block $\B$ to the full lattice.

\subsection 5.1 Acceptance probability

Following the general strategy summarized in 
subsect.~A.5, the probability 
distribution $P_{\B}[U]$ is considered to be an approximation to
the exact distribution $P[U]$.
An auxiliary pseudo-fermion field
$\chi$ is then added to the system with action
\equation{
  \Schi=\left(\chi,(\QB+\QBs)Q^{-1}\rhoB^2Q^{-1}(\QB+\QBs)\chi\right),
  \enum
}
where a local shape function
\equation{
  \rhoB(x)=\PrdB(x)+\epsilon\left(1-\PrdB(x)\right),
  \enum
}
with parameter $\epsilon>0$,
has been included for reasons that will become clear in a moment.
The simultaneous probability distribution of the enlarged theory,
\equation{
  \hat{P}[\chi,U]={1\over\hat{\cal Z}}\,
  \rme^{-\Schi}P_{\B}[U],
  \enum
}
reproduces the correct distribution when the auxiliary field is integrated
out. Note that $\rhoB$ is invertible and independent of the fields.
The associated determinant is therefore a constant that cancels 
out once the normalization factors are taken into account.

Before the acceptance--rejection step can be applied,
a pseudo-fermion 
field $\chi$ must be generated randomly with conditional probability
$\hat{P}[\chi\kern1pt|\kern1pt U]$. 
This is easily achieved by setting 
\equation{
  \chi=\left(\QB+\QBs\right)^{-1}Q\rhoB^{-1}\eta,
  \enum
}
where $\eta$ is chosen with probability proportional 
to $\rme^{-(\eta,\eta)}$.
The proposed field $U'$ generated by the block simulation algorithm
is then accepted with probability
\equation{
  P_{\rm acc}=\min\left\{1,\rme^{-\Delta\Schi}\right\},
  \qquad
  \Delta\Schi\equiv\left.\Schi\right|_{U\to U'}-\Schi,
  \enum
}
which ensures that the gauge field is correctly updated
with a transition probability that satisfies detailed balance
with respect to the exact distribution $P[U]$.

\subsection 5.2 Elimination of the volume terms

In the following lines it is shown that $\Delta\Schi$
only depends on the values of $\eta$ on the exterior boundary $\dB$,
up to terms that are proportional to the parameter $\epsilon$
of the shape function $\rhoB$ and that can, therefore, be eliminated
by taking the limit $\epsilon\to0$.

The proof starts from the identities
\equation{
  \Delta\Schi=\left(\zeta,\zeta\right)-\left(\eta,\eta\right),
  \enum
  \next{2.5ex}
  \zeta\equiv\rhoB Q'^{-1}\left(\QB'+\QBs'\right)
  \left(\QB+\QBs\right)^{-1}Q
  \rhoB^{-1}\eta
  \noenum
  \next{2ex}
  {\phantom{\zeta}}=\eta+
  \rhoB Q'^{-1}\left(\QB'-\QB\right)
  \left(\QB+\QBs\right)^{-1}\left(\QdB+\QdBs\right)\rhoB^{-1}\eta,
  \enum
}
where the last line has been obtained using
the block decomposition (3.2) of the Dirac operator
and the fact that $\QBs$, $\QdB$ and $\QdBs$ are independent of the 
active link variables.
The crucial observation is then that
all terms proportional to $\QdBs\rhoB^{-1}\eta$ vanish
since $\QB'-\QB$ acts in the 
subspace of fermion fields supported on $\B$.
Recalling eqs.~(3.5) and (5.2), this leads to 
\equation{
  \zeta=\eta+
  \rhoB Q'^{-1}\left(\QB'-\QB\right)\QB^{-1}\QdB\eta
  \enum
}
and thus to an expression
\equation{
  \Delta\Schi=2\Re\kern-1pt
  \left(\PrdB\eta,\xi\right)+\left(\xi,\xi\right)
  +\rmO(\epsilon),
  \enum   
  \next{2ex}
  \xi\equiv
  \PrdB Q'^{-1}\left(\QB'-\QB\right)\QB^{-1}\QdB\eta,
  \enum
}
of the type announced above. 

At this point the parameter $\epsilon$ 
can safely be taken to zero because all terms that are inversely
proportional to $\epsilon$ have disappeared.
Moreover, since only 
the component $\PrdB\eta$ of the
field $\eta$ enters the 
final expressions, the constraint
\equation{
  \eta=\PrdB\eta
  \enum
}
may now be imposed without loss. The field 
is thus restricted to the linear space of boundary values for 
the Dirichlet problem on $\B$ (cf.~subsect.~3.3).

\subsection 5.3 Acceptance rate

Evidently the whole approach will fail in practice unless
the acceptance probability (5.5) is reasonably large on average.
This will obviously be the case if
$\Delta\Schi$ is only rarely greater than $1$,
and the question thus arises of whether there is any theoretical
reason to expect this to be so.

Some insight into the problem can be gained by assuming, for a moment,
that $U'$ differs from $U$ only on a single link $l$ in $\B$.
In eq.~(5.9) the first term then
involves the propagation of the random field
$\eta$ from the exterior boundary $\dB$ to the endpoints of $l$
and a second propagation from there back to the boundary.
The other term has a similar structure
but involves altogether four quark propagators.

Quark propagators 
typically decay like
$d^{-3}$ in the short-distance regime and more
rapidly at larger distances $d$.
The contributions to $\Delta\Schi$
with two quark propagators
are thus suppressed by a factor $d^{-3}d'^{-3}$ at least, 
where $d$ and $d'$ are the distances of the link $l$
from the initial and final points on $\dB$.
The sum over all these points then still needs to be performed,
but since $\eta$ is a random field
there are strong cancellations in this sum and on average
the result is enhanced by a factor proportional to
the number of boundary points only (rather than its square).
An important suppression factor thus remains, particularly when
the link $l$ is a fair distance away from the boundary.

Reasonable acceptance rates
can thus be expected if the proposed changes in the link variables
on the links closer to the boundary of the block are
not too large. By adjusting the speed factor $\gamma(x,\mu)$ in the
molecular dynamics equations (4.3),(4.4), this can be easily achieved.
Note that there must also be important cancellations in the sum over
the active link variables that is implicit in eq.~(5.10),
because they change in random directions in group space.

\subsection 5.4 Numerical exercises

Some numerical confirmation of this qualitative argumentation
is clearly desirable at this point. 
The decay of the quark propagator away from
the boundary of the block, for example, can be checked 
by calculating the 
solution $\psi=-\QB^{-1}\QdB\eta$ of the Dirichlet problem on $\B$.
A representative result of such studies 
for the average of $\left|\psi(x)\right|^2$
over all boundary values and gauge fields
is reported in fig.~4.

Another check on the correctness of the theoretical argumentation 
is obtained by computing the 
average of the acceptance probability $P_{\rm acc}$
in the quenched approximation, where the proposed changes in 
the active link variables 
are generated using a link update algorithm.
The results in this case are rather encouraging too
and they demonstrate, in particular, the importance of the 
elimination of the volume terms.

\topinsert
\vbox{
\vskip0.0cm
\epsfxsize=8.0cm\leftline{\hfill\epsfbox{psi.eps}\hfill}
\vskip0.4cm
\figurecaption{%
Average magnitude squared of the solution $\psi(x)$
of the Dirichlet problem (3.6),(3.7) on a $12^4$ block
as a function of the distance $d$ from the exterior boundary
of the block. The data shown are from a simulation of quenched QCD
at a lattice spacing of about $0.1$ fm.
Diamonds, squares and circles
correspond to values of the quark mass
where the pion mass is known to be
approximately equal to $320$, $457$ and $579$ MeV respectively [\ref{CPPACS}].
}
\vskip0.1cm
}
\endinsert

In all these studies the
dependence of the calculated quantities on the quark mass
appears to be fairly weak, as long as the edges of the block
are smaller than $1$~fm or so.
This is perhaps not totally surprising in view of what has been said above,
but the observation suggests
that the algorithm will remain effective also at small quark masses.

\subsection 5.5 Hierarchical filter

As was briefly discussed in subsect.~2.3,
a hierarchical structure should be adopted on large lattices
where the proposed gauge field configurations are filtered through 
a sequence of blocks of increasing size.
The filter involves an acceptance--rejection step
for each transition from a block $\B$ to the next larger
block $\Omega$.
Equations (5.9)--(5.11) remain valid
in this case if $Q'$ is replaced by $Q'_{\Omega}$.
An important point to note here is that the 
distance of the links where the active link variables reside
to the boundary of the blocks
is increasing. The acceptance probability
for the step from one block to the next is therefore expected to 
rise quickly. 

The proposed configurations that remain after the application of the 
hierarchical filter must finally be submitted to a global
acceptance--rejection step. 
If several configurations are submitted simultaneously, as suggested in 
subsect.~2.3, the acceptance probability is again given by 
eqs.~(5.5) and (5.9)--(5.11), where $\B$ now stands for the union 
of the big blocks that contain the
proposed configurations. The boundary $\dB$ is then the 
union of the boundaries of these blocks and $\PrdB$ the sum of the 
corresponding characteristic functions. In particular,
there are terms in $\Delta\Schi$ that couple the random field
$\eta$ on the boundaries of different blocks.

\section 6. Miscellaneous remarks

\vskip0ex\noindent
(1)~{\it Preconditioning}.
As is generally the case in lattice QCD, the preconditioning of 
the Dirac operator can be expected to have a significant impact on the 
performance of the proposed algorithm. In particular, 
it is straightforward to implement even--odd 
preconditioning both in the 
block simulation algorithm and in the global acceptance--rejection step.

\vskip1ex\noindent
(2)~{\it Block simulation algorithm}.
There is a range of algorithms that can be used 
to generate the proposed gauge field configurations on a given block.
In particular,
most improvements of the HMC and the PHMC algorithms
that have been introduced over the years,
such as the two-boson method [\ref{HasenbuschTB},\ref{HasenbuschJansen}], 
should be beneficial here too.

A more radical possibility is to generate the configurations through a
standard link update algorithm,
using an adapted gauge field action, and to apply a stochastic 
acceptance--rejection
step to correct for the quark determinant 
[\ref{Hasenbusch}--\ref{PSD}]. 
Following the lines of subsect.~5.2,
the random pseudo-fermion field that needs 
to be introduced in the correction step can be reduced to the support
of $\QB'-\QB$ in this case. This will no doubt increase
the average acceptance probability, but whether the approach can compete
with the block HMC algorithm, for example, remains to be seen.

\vskip1ex\noindent
(3)~{\it Parallel processing}.
The proposed algorithm is well suited for parallel computers, because 
distant blocks can be updated independently of each other.
Different distributions of the workload are conceivable, 
and it is not required that the smallest blocks be 
processed on single nodes, although in this case 
the communication overhead is minimized.
It is, incidentally, advisable to translate
the gauge field after each cycle rather than 
shifting the block positions. The program then always operates
on the same blocks and their position can be chosen so as to 
fit the processor grid in the best possible way.

\section 7. Conclusions

The problem to find efficient simulation algorithms
for unquenched lattice QCD in the chiral regime has been around 
for many years.
Whether the approach described in this paper provides a viable
solution is not certain and can only be decided 
after extensive numerical tests have been performed.

It is quite clear, however, that significant progress in this 
area can only be made if the short- and long-distance
effects of the sea quarks are treated differently.
This insight is not new and has motivated the 
truncated determinant approximation of 
Duncan et al.~[\ref{DuncanI},\ref{DuncanII}], for example,
and a modification of the HMC algorithm with multiple
molecular dynamics time scales [\ref{PeardonTS}].
To some extent, the two-boson method studied in 
refs.~[\ref{HasenbuschTB},\ref{HasenbuschJansen}]
can also be put under this general heading.

The algorithm proposed here separates
the short-distance effects from the long-distance ones
by updating the gauge field on physically small 
blocks of lattice points that are visited sequentially.
On the blocks the theory
can be simulated even at vanishing quark masses,  
because the chosen boundary conditions imply an infrared cutoff
on the spectrum of the Dirac operator.
The long-distance effects of the sea quarks
are then incorporated by a sequence of acceptance--rejection
steps that lead from the small blocks to larger blocks and eventually
to the full lattice.

\vskip1ex
I wish to thank Martin Hasenbusch and Ulli Wolff for correspondence on 
fermion simulation algorithms and Peter Weisz for a critical reading
of the manuscript.
The computer time for the numerical studies reported in sect.~5
has been kindly provided by DESY--Hamburg 
and by the Institute for Theoretical Physics at the University of Berne.

\appendix A. Transition probabilities 

To prove the correctness of 
the algorithm proposed in this paper
it suffices to write down the transition probabilities that 
are associated to the various algorithmic steps
and to recall some basic facts from the general theory of numerical 
simulations. For completeness, the relevant theoretical
results are summarized in this appendix.

\subsection A.1 Convergence theorem

To avoid any unnecessary complications, 
an abstract discrete system will be considered 
with a finite number of states $s$.
The thermal equilibrium distribution of the states is denoted
by $P(s)$ and it is assumed that $P(s)>0$ for all~$s$.

Numerical simulation algorithms for this system (as far as they 
follow the standard procedures) are defined
by a transition probability $T(s\to s')$ that satisfies

\vskip2.0ex
\halign{\hskip1.0em#\hfil&\hskip0.5em%
\vtop{\parindent=0pt\hsize=11.5cm\strut#\strut}\cr
$1.$&%
$T(s\to s')$ is  non-negative and such that
$\sum_{s'}\,T(s\to s')=1$ for all $s$.
\cr
\noalign{\vskip1.5ex}
$2.$&%
The equilibrium distribution is preserved, i.e.~for
all states $s'$ the equation $\sum_sP(s)T(s\to s')=P(s')$ holds.
\cr
\noalign{\vskip1.5ex}
$3.$&%
The recurrence probability $T(s\to s)$ is non-zero for all $s$ and
any state can be reached from any other state in a finite number of 
transitions.
\cr
}

\vskip1.5ex
\noindent
In practice the simulation starts
from an initial state $s_0$ and generates a sequence of states
$s_1,s_2,\ldots$ recursively with transition probability
$T(s_k\to s_{k+1})$.
The fundamental theorem of simulation theory then asserts that
the states in the sequence will be asymptotically distributed
according to the equilibrium distribution.

\subsection A.2 Composition rule \& detailed balance

If $T_1$ and $T_2$ are any two transition probabilities
that satisfy conditions 1--3, it is trivial to show that
the same is true for the composed transition probability
\equation{
  T(s\to s')=\sum_rT_1(s\to r)T_2(r\to s').
  \enum
}
The associated algorithm generates an intermediate state $r$
with probability $T_1(s\to r)$ and then the new state $s'$
with probability $T_2(r\to s')$. Evidently any number of 
algorithms can be combined in this way.

Transition amplitudes that satisfy
detailed balance,
\equation{
  P(s)T(s\to s')=P(s')T(s'\to s),
  \enum
}
are an important special case. If detailed balance holds, property 2
is an immediate consequence of property 1.
However, the composed transition probability (A.1) in general does not satisfy 
detailed balance if $T_1$ and $T_2$ do, because the order of 
the factors in eq.~(A.1) matters. 
It is possible to correct for this deficit by
choosing the order randomly or by forming reversion-symmetric products.

\subsection A.3 Acceptance--rejection method

Valid algorithms may often be obtained 
from transition amplitudes
$T_0(s\to s')$ that satisfy detailed balance with respect to
some approximate distribution $P_0(s)$ [\ref{Metropolis}].
The idea is to first propose 
a new state $s'$ according to this probability 
and to accept it with probability
\equation{
  P_{\rm acc}(s,s')=\min\left\{1,{P_0(s)P(s')\over P(s)P_0(s')}\right\}.
  \enum
}
Otherwise (i.e.~when $s'$ is rejected) 
the old state $s$ is taken to be the new one.

The transition amplitude that is associated to this algorithm,
\equation{
  T(s\to s')=T_0(s\to s')P_{\rm acc}(s,s')+
             \delta_{ss'}-\delta_{ss'}
             \sum_rT_0(s\to r)P_{\rm acc}(s,r),
  \enum
}
satisfies detailed balance and also conditions 1 and 3.
It must be stressed, however, that for this statement 
to be true it is essential that $T_0(s\to s')$
satisfies detailed balance with respect to the distribution $P_0(s)$
and not only condition 2.

\subsection A.4 Using auxiliary stochastic variables

When the equilibrium probability distribution is 
too complicated to be treated directly, an equivalent but more
accessible
system may sometimes be found that involves auxiliary stochastic variables.
In the case of the HMC algorithm, 
for example, the auxiliary variables 
are the pseudo-fermion field and the
momenta of the link variables [\ref{HMC}]. 
A simulation algorithm may
then be set up for the enlarged system, and this
leads to a correct algorithm for the original system under certain
conditions.

In abstract terms the states of the enlarged system are labelled by
pairs $v,s$, where $v$ and $s$ represent the auxiliary and the 
basic variables
respectively. The equi\-valence to the original system is then
expressed through the identity
\equation{
   P(s)=\sum_v\hat{P}(v,s)
   \enum
}
in which $\hat{P}(v,s)$ denotes the equilibrium distribution of the 
enlarged
system. In the following the conditional probability
\equation{
   \hat{P}(v|s)=\hat{P}(v,s)/P(s)
   \enum
}
to find $v$ given $s$ will play an important r\^ole.

Now if $\hat{T}(v,s\to v',s')$ is any given transition probability 
for the enlarged system
that satisfies conditions $1$ and $2$ (but not necessarily condition 3),
an update algorithm for the original system is obtained as follows.
Starting from an initial state $s$,
the auxiliary variables $v$ are first chosen randomly
with conditional probability $\hat{P}(v|s)$.
After that, a state $v',s'$ is
generated with probability $\hat{T}(v,s\to v',s')$ and $s'$ is taken
to be the new state of the original system.
It is straightforward to show that 
the associated transition probability
\equation{
  T(s\to s')=\sum_{v,v'}\hat{P}(v|s)\hat{T}(v,s\to v',s')
  \enum
}
satisfies conditions 1 and 2. 
Moreover, detailed balance holds if the transition probability 
$\hat{T}(v,s\to v',s')$ has this property relative
to the equilibrium distribution of the enlarged system.
Whether condition 3 is fulfilled depends on 
both $\hat{P}(v|s)$ and $\hat{T}(v,s\to v',s')$
and has to be verified in each case.

\subsection A.5 Stochastic acceptance--rejection method

The starting point here is again an algorithm
with transition probability $T_0(s\to s')$ 
that satisfies detailed balance with respect to
some approximate distribution $P_0(s)$.
An enlarged system is then considered, exactly
as above, but the auxiliary variables
are now only used in the acceptance--rejection step where
(A.3) is replaced by
\equation{
   P_{\rm acc}(s,s')=\sum_v\hat{P}(v|s)\min\left\{
   1,{P_0(s)\hat{P}(v,s')\over\hat{P}(v,s)P_0(s')}\right\}.
   \enum
}
This means that once a new state
$s'$ of the original system is proposed, 
a random choice of the auxiliary variables $v$
first needs to be made, with conditional probability
$\hat{P}(v|s)$, before it can be decided whether $s'$ should
be accepted or not.

The correctness of this method follows from
the identity
\equation{
  \left[P(s)/P_0(s)\right]P_{\rm acc}(s,s')
  =
  \left[P(s')/P_0(s')\right]P_{\rm acc}(s',s)
  \enum
}
that is easily derived from the definition (A.8)
and the properties of the distributions 
$P_0(s)$ and $\hat{P}(v,s)$.
In particular, the associated transition probability (A.4) 
satisfies detailed balance.
It has recently been noted, however, that [\ref{PSD}] 
\equation{
  P_{\rm acc}(s,s')\leq\min\left\{\sum_v\hat{P}(v|s),
  \sum_v{P_0(s)\hat{P}(v,s')\over P(s)P_0(s')}\right\}
  \noenum
  \next{2.0ex}
  {\phantom{P_{\rm acc}(s,s')}}
  =\min\left\{1,{P_0(s)P(s')\over P(s)P_0(s')}\right\},
  \enum
}
which shows that the stochastic acceptance--rejection method tends to be
less efficient than the non-stochastic method. 

\beginbibliography


\bibitem{HMC}
S. Duane, A. D. Kennedy, B. J. Pendleton, D. Roweth,
Phys. Lett. B195 (1987) 216


\bibitem{PHMCI}
P. de Forcrand, T. Takaishi,
Nucl. Phys. B (Proc. Suppl.) 53 (1997) 968

\bibitem{PHMCII}
R. Frezzotti, K. Jansen,
Phys. Lett. B 402 (1997) 328;
Nucl. Phys. B 555 (1999) 395 and 432


\bibitem{MB}
M. L\"uscher, Nucl. Phys. B418 (1994) 637


\bibitem{Schwarz}
H. A. Schwarz, Gesammelte Mathematische Abhandlungen, vol. 2
(Springer Verlag, Berlin, 1890)

\bibitem{Saad}
Y. Saad, Iterative methods for sparse linear systems 
(PWS Publishing Company, Boston, 1996); see also
{\tt http://www-users.cs.umn.edu/\~{}saad/}


\bibitem{Hasenbusch}
M. Hasenbusch,
Phys. Rev. D59 (1999) 054505


\bibitem{FrezzottiEtAl}
R. Frezzotti et al. (ALPHA collab.),
Comput. Phys. Commun. 136 (2001) 1


\bibitem{HasenfratzKnechtli}
A. Hasenfratz, F. Knechtli,
Comput. Phys. Commun. 148 (2002) 81

\bibitem{AlexandruHasenfratz}
A. Alexandru, A. Hasenfratz,
Phys. Rev. D65 (2002) 114506;
{\it ibid.} D66 (2002) 094502


\bibitem{PSD}
F. Knechtli, U. Wolff,
Dynamical fermions as a global correction,
hep-lat/0303001


\bibitem{Wilson}
K. G. Wilson, Phys. Rev. D10 (1974) 2445

\bibitem{SW}
B. Sheikholeslami, R. Wohlert,
Nucl. Phys. B 259 (1985) 572

\bibitem{OaImp}
M. L\"uscher, S. Sint, R. Sommer, P. Weisz,
Nucl. Phys. B478 (1996) 365


\bibitem{SchroedingerGauge}
M. L\"uscher, R. Narayanan, P. Weisz, U. Wolff,
Nucl. Phys. B384 (1992) 168

\bibitem{SchroedingerFermion}
S. Sint,
Nucl. Phys. B421 (1994) 135

\bibitem{SchroedingerFullQCD}
K. Jansen, R. Sommer (ALPHA collab.),
Nucl. Phys. B530 (1998) 185 [E: {\it ibid.}\/ B643 (2002) 517]


\bibitem{LowMode}
L. Giusti, C. Hoelbling, M. L\"uscher, H. Wittig,
Numerical techniques for lattice QCD in the $\epsilon$-regime,
hep-lat/0212012, to appear in Comput. Phys. Commun.


\bibitem{DaviesEtAl}
C. T. H. Davies et al.,
Phys. Rev. D41 (1990) 1953


\bibitem{CPPACS}
S. Aoki et al. (CP-PACS collab.),
Phys. Rev. D67 (2003) 034503


\bibitem{HasenbuschTB}
M. Hasenbusch,
Phys. Lett. B519 (2001) 177

\bibitem{HasenbuschJansen}    
M. Hasenbusch, K. Jansen,     
Speeding up lattice QCD simulations with clover improved Wilson fermions,
hep-lat/0211042


\bibitem{DuncanI}
A. Duncan, E. Eichten, H. Thacker,
Phys. Rev. D59 (1999) 014505

\bibitem{DuncanII}
A. Duncan, E. Eichten, Y. Yoo,
Unquenched QCD with light quarks,
hep-lat/0209123


\bibitem{PeardonTS}
M. Peardon, J. Sexton (TrinLat Collab.),
Multiple molecular dynamics time scales in hybrid Monte Carlo fermion
simulations,
hep-lat/0209037


\bibitem{Metropolis}
N. Metropolis et al.,
J. Chem. Phys. 21 (1953) 1087

\endbibliography

\bye